\documentclass[lettersize,journal]{IEEEtran}
\usepackage{amsmath,amsfonts}
\usepackage{algorithmic}
\usepackage{algorithm}
\usepackage{array}
\usepackage{textcomp}
\usepackage{stfloats}
\usepackage{url}
\usepackage{verbatim}
\usepackage{graphicx}
\usepackage{cite}
\usepackage{xcolor}
\usepackage{subcaption}
\begin{document}

\title{Multipath TCP with Single Radio Access Technologies: a Paradox or an Opportunity?}

\author{João Torrinhas, Miguel Luís, Duarte Dias, Pedro Rito, Susana Sargento
    \thanks{All authors are with Instituto de Telecomuniações, Campus Universitário de Santiago, 3810-193 Aveiro, Portugal. J. Torrinhas, P. Rito and S. Sargento are also with Departamento de Electrónica, Telecomunicações e Informática, Universidade de Aveiro, 3810-193 Aveiro, Portugal. M. Luís is also with Instituto Superior Técnico, Universidade de Lisboa, Av. Rovisco Pais 1, 1049-001 Lisboa, Portugal. E-mails: \{joao.torrinhas,duarterochadias,pedrorito,susana\}@ua.pt, nmal@av.it.pt.}
    \vspace{-7mm}
}



\maketitle

\begin{abstract}
This paper addresses the use of Multipath Transmission Control Protocol (MPTCP) in a single Radio Access Technology (RAT) network. Different from other studies where multiple RATs are explored by the MPTCP, a situation that cannot be always guaranteed, due to lack of coverage for example, in this work we assess and evaluate the capability of MPTCP to operate over a single RAT environment. With a vehicular network as use case, we show how the IEEE 802.11p interface is shared among the multiple logical links created between the On-Board Unit (OBU) and the several Road Side Units (RSUs) in its range, supporting the different MPTCP subflows. The results, obtained through experimentation with real vehicular networking hardware, show that MPTCP allows for seamless handovers, ensuring continuous, stable and efficient communication in highly mobile environments. 
\end{abstract}


\section{Introduction}\label{sec::1}
Multipath TCP (MPTCP) is an enhanced version of Transmission Control Protocol (TCP), designed to increase application performance and reliability by allowing concurrent use of multiple paths between the source and the destination~\cite{bib:rfc8684}. A fundamental aspect of MPTCP is its resilience in scenarios of heterogeneous access technologies, exploiting the diversity of access technologies, dynamically distributing data across multiple paths. This ensures efficient and robust communication, allowing the protocol to adapt to the various conditions and capabilities of each network. This protocol also ensures seamless handovers between various network interfaces, such as Wi-Fi and cellular networks, a feature particularly beneficial in mobile environments, characterized by frequent changes in the network topology. 

Vehicular \textit{ad hoc} Networks (VANETs) are advanced systems built on the principles of Mobile \textit{ad hoc} Networks. They are designed to support the communication between vehicles and with the roadside infrastructure, providing services such as safety warnings, making them a key part in the Intelligent Transportation Systems (ITS) framework. Taking advantage of the use of wireless communication technologies, VANETs allow the creation of real-time connections between vehicles, pedestrians, roadside infrastructure and sensors, improving the safety of road entities and increasing the efficiency of the transport ecosystem. Due to the high mobility of its nodes --- namely vehicles, equipped with On-Board Units (OBUs) ---, it is essential to maintain a stable and reliable connection with Road Side Units (RSUs) to ensure continuous communication flow and prevent the loss of critical road safety information. To achieve this, seamless handovers are necessary, meaning a vehicle must transition its connection from one RSU to another as it moves along its route without perceivable interruptions or service disruptions.

IEEE 802.11p has been assumed as the primary radio access technology in vehicular environments. This standard, derived from the IEEE 802.11a standard, and recently replaced by a new amendment (IEEE 802.11bd), was designed to better cope with the mobility of its elements, namely their velocity, therefore abolishing the association and authentication procedures, among other features~\cite{bib:80211p}. Despite its long communication range, which can, in some situations, achieve up to 1 km~\cite{bib:nxp_cits}, OBUs can also be equipped with cellular radio interfaces, 4G and 5G, to complement and increase the connectivity of mobile elements. This configuration of multiple Radio Access Technologies (RAT) in VANETs has been explored by the scientific community to implement and evaluate the MPTCP protocol \cite{bib:Mena17, bib:Liu23, bib:abed24}.

This work presents a different study on the suitability and performance of MPTCP in VANETs. First, we consider the scenario where, temporarily or permanently, a single-RAT is available for the communication between OBUs and RSUs, for example, due to implementation decisions or lack of connectivity on the second RAT. Second, we consider a vehicular network architecture, enabled by Software Defined Network (SDN), where, even in a single-RAT situation, an OBU is capable of creating simultaneous logical links with different RSUs in its communication range. This characteristic allows MPTCP to operate even in single-RAT scenarios. To better understand the novelty of this study, note that operating MPTCP in this single-RAT context differs from using \textit{traditional} TCP in a single-RAT scenario, as MPTCP leverages multiple distinct paths between OBUs and RSUs within its range, unlike traditional TCP, which relies on a single path.

The results, obtained through experiments using real vehicular networking hardware (OBUs and RSUs) with emulated mobility, show that MPTCP can be successfully used in single-RAT environments where the radio resources are shared between the several active logical links, which are then explored by the different MPTCP subflows. To the best of our knowledge, this is the first time that MPTCP is studied in single-RAT scenarios, which can be considered as a paradox given the nature of such transport protocol.


The remainder of this paper is as follows. Section~\ref{sec::2} introduces the SDVN used as overlay network to manage the connectivity of the vehicular network. Section~\ref{sec::3} explains how MPTCP was integrated in the SDVN solution. Finally, Section~\ref{sec::4} presents the performance results and Section~\ref{sec::5} concludes the paper.

\section{SDN-based Vehicular Network}\label{sec::2}

The MPTCP protocol operates at the transport layer, allowing simultaneous transmission of traffic across MPTCP subflows. In a VANET scenario, if the OBU is communicating with the infrastructure, each MPTCP subflow is created over a logical link between the OBU and a given RSU via IEEE 802.11p. Thus, an underlying network is required to manage the connectivity between OBUs and RSUs, resulting in the creation and removal of logical links, as well as in the management of traffic between the RSUs and the server. 

\begin{figure}[ht]
    \begin{center}
    \includegraphics[width=0.33\textwidth]{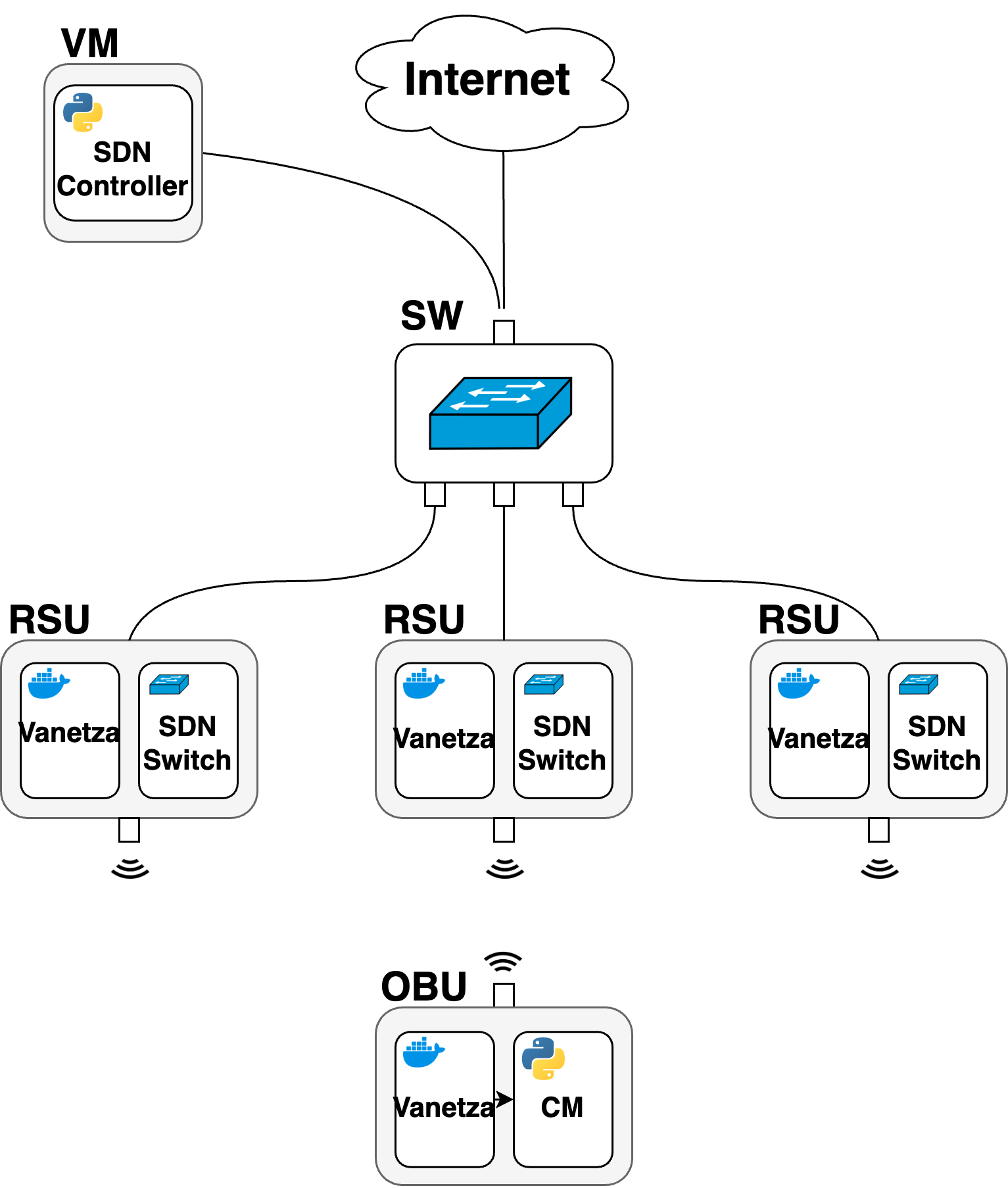}
    \caption{Software-Defined Vehicular Network architecture: single-RAT is assumed, with IEEE 802.11p/ITS-G5 vehicular communication stack.}
    \label{fig:sdvn_arch}
    \end{center}
\end{figure}

The explored SDVN, detailed in \cite{bib:sdvn_cits2024} and simplified in Fig.~\ref{fig:sdvn_arch}, consists of OBUs, representing the mobile nodes --- the vehicles --- where a single RAT is available, namely IEEE 802.11p with an ITS-G5 vehicular stack, and RSUs, acting as gateways to extend the connectivity of the OBUs, namely, to the Internet. The static infrastructure operates within an SDN domain, where each RSU operates as an SDN switch and an SDN controller which oversees the flow management, ensuring that downlink traffic is properly redirected to the correct logical links between the RSU(s) and the different OBUs. The management done by the SDN controller relies on Cooperative Intelligent Transport System (C-ITS) messages exchanged between RSUs and OBUs, through the vehicular network stack Vanetza \cite{bib:Rosmaninho24}, providing positioning information, signal strength, and other details about each mobile node to the SDN controller. A similar management must be performed by the OBU with respect to the uplink traffic, under the responsibility of the Connection Manager (CM). 

\section{MPTCP in Single-RAT Vehicular Networks}\label{sec::3}

\begin{figure}[ht]
    \begin{center}
    \includegraphics[width=0.33\textwidth]{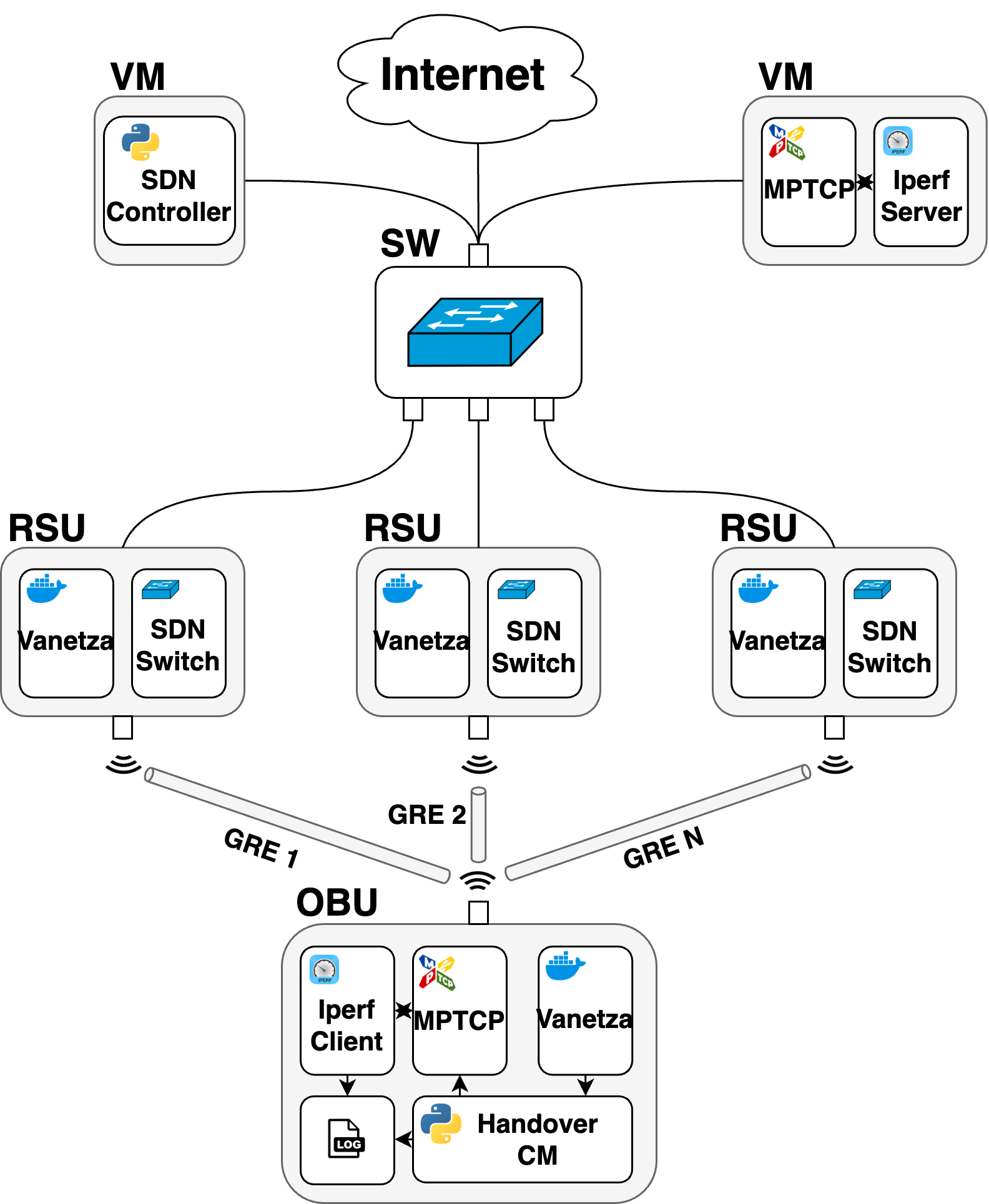}
    \caption{Integration of MPTCP in the SDVN architecture.}
    \label{fig:Architecture_With_MPTCP}
    \end{center}
\end{figure}

To fully explore the operation of MPTCP, the vehicular network must support the simultaneous transmission of IP traffic through different logical links between one OBU and the different RSUs in its range. One solution lies on the use of GRE tunnels in the wireless link between the OBU and the RSU, creating an overlay network\footnote{It is important to highlight that this work does not aim to evaluate the performance of such overlay solution, nor the SDVN. Instead, it provides the required configuration for the proper implementation of MPTCP subflows over the several wireless logical links between the OBU and the RSUs. We are aware that other tunneling configurations could be used, such as VXLANs.}~\cite{bib:Ferreira24}. In this way, the OBU's wireless radio interface is shared by multiple GRE tunnels, associated with several RSUs, on which MPTCP subflows will be created and removed, as illustrated in Fig.~\ref{fig:Architecture_With_MPTCP}.

Through the use of multiple logical links (supported by the GRE tunnels) and MPTCP capabilities, data transmission is facilitated due to the use of multiple paths between the OBU, hosting the MPTCP client, and the Virtual Machine entity, hosting the MPTCP server. Such configuration allows the OBU to perform multiple handovers between RSUs, transmitting data simultaneously through the MPTCP subflows associated with these RSUs, ensuring a stable and continuous MPTCP connection along its movement.

In this configuration, the CM must extend its capabilities to also manage the MPTCP subflows. According to the information provided by the C-ITS messages (location, signal strength, etc.), the CM, here denoted as Handover CM, is responsible for the creation and removal of MPTCP subflows over pre-established GRE tunnels. Such task is needed because MPTCP is not capable of automatically create and remove subflows while the OBU is traveling. Operating at transport layer, MPTCP is not aware of connectivity changes, \textit{i.e.}: when an OBU handovers to another RSU, the MPTCP protocol does not automatically start sending data along the new path unless a new subflow is manually created over such path. 

On the other hand, MPTCP handles the resilience part well: if one path fails, the protocol redirects data to other available paths that are still operational. Several MPTCP packet scheduling algorithms/policies exist, ranging from prioritizing paths with the highest available bandwidth to paths with the lowest Round Trip Time (RTT). In network configurations with multiple RATs, it is reasonable to assume that one access technology presents higher bandwidth, or lower RTT, than the others, leading the MPTCP to prefer one path over the others. However, in single-RAT scenarios, as the one in evaluation in this work, it is not clear how the characteristics of a given logical link are different from the others, since the same radio access technology is shared by all the active logical links.  

\section{Performance Evaluation}\label{sec::4}

The MPTCP version used in this study is MPTCPv1\footnote{Multipath TCP Project available at \url{https://www.mptcp.dev/}.}, which is the latest version and, at the time, the only one supported by the upstream Linux kernel. Furthermore, the default MPTCP scheduling algorithm (minRTT) is used -- where data are first transmitted in the subflow with the lowest RTT until its congestion window is full, and then it starts sending data on the subflow with the next highest RTT, and so on --, along with the Linked Increase Algorithm coupled congestion control algorithm, which couples the behavior of multiple paths~\cite{rfc6356}.

The network setup used in the performance evaluation follows the architecture presented in Section~\ref{sec::2}. One OBU, represented by a PC Engines APU3 platform and equipped with an IEEE 802.11p and the ITS-G5 vehicular communication stack, was configured with the aforementioned MPTCP framework and an iPerf3 client\footnote{iPerf3 available at https://iperf.fr/.}. To assess the MPTCP behavior in the presence of multiple wireless logical links, three RSUs were also configured, using the same hardware as the OBU, but with an additional Ethernet interface to support the connection with the Internet. Finally, a VM entity was configured with the aforementioned MPTCP implementation and an iPerf3 server.

\begin{figure}[ht]
    \begin{center}
    \includegraphics[width=0.35\textwidth]{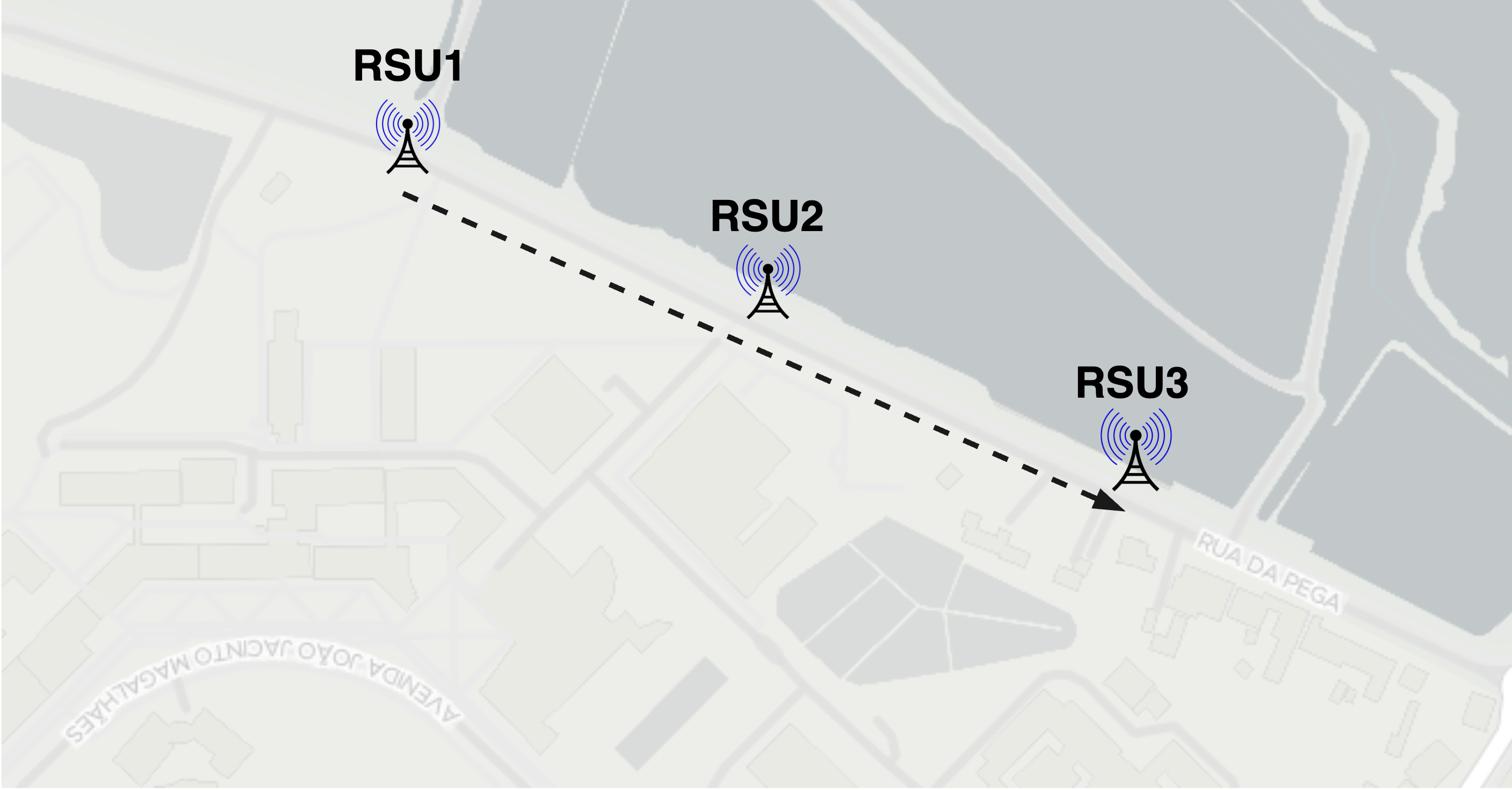}
    \caption{OBU's emulated path.}
    \label{fig:Traject_Lab}
    \end{center}
    \vspace{-5mm}
\end{figure}

Although we have used a real communication vehicular network, the mobility of the OBU was emulated, meaning that the existence of a connection between the OBU and the different RSUs was done in controlled environment --- turning the RSUs on and off. The OBU's emulated path, following a straight line, is displayed in Fig.~\ref{fig:Traject_Lab}: at $t=0$ s, the OBU is in communication range with a single RSU, RSU1; then, at $t=20$ s, the OBU becomes in range of RSU2, creating two logical communication links with RSU1 and RSU2, a situation that remains until $t=50$ s; until $t=100$ s, the OBU is in communication range of a single RSU, RSU2, and from that moment on until the end of the experimentation, the OBU is in communication range of two RSUs, RSU2 and RSU3.

Finally, to better assess the behavior and performance of the MPTCP protocol, two scenarios were considered: one with no manipulation on the network characteristics, and a second one where an additional delay, 200 ms, was applied to the wireless radio interface of the second RSU, affecting the logical link between the OBU and this RSU, and therefore, the MPTCP involving these two nodes.

\begin{figure}[ht]
    \centering
    \begin{subfigure}[b]{0.4\textwidth}
        \centering
        \includegraphics[width=\textwidth]{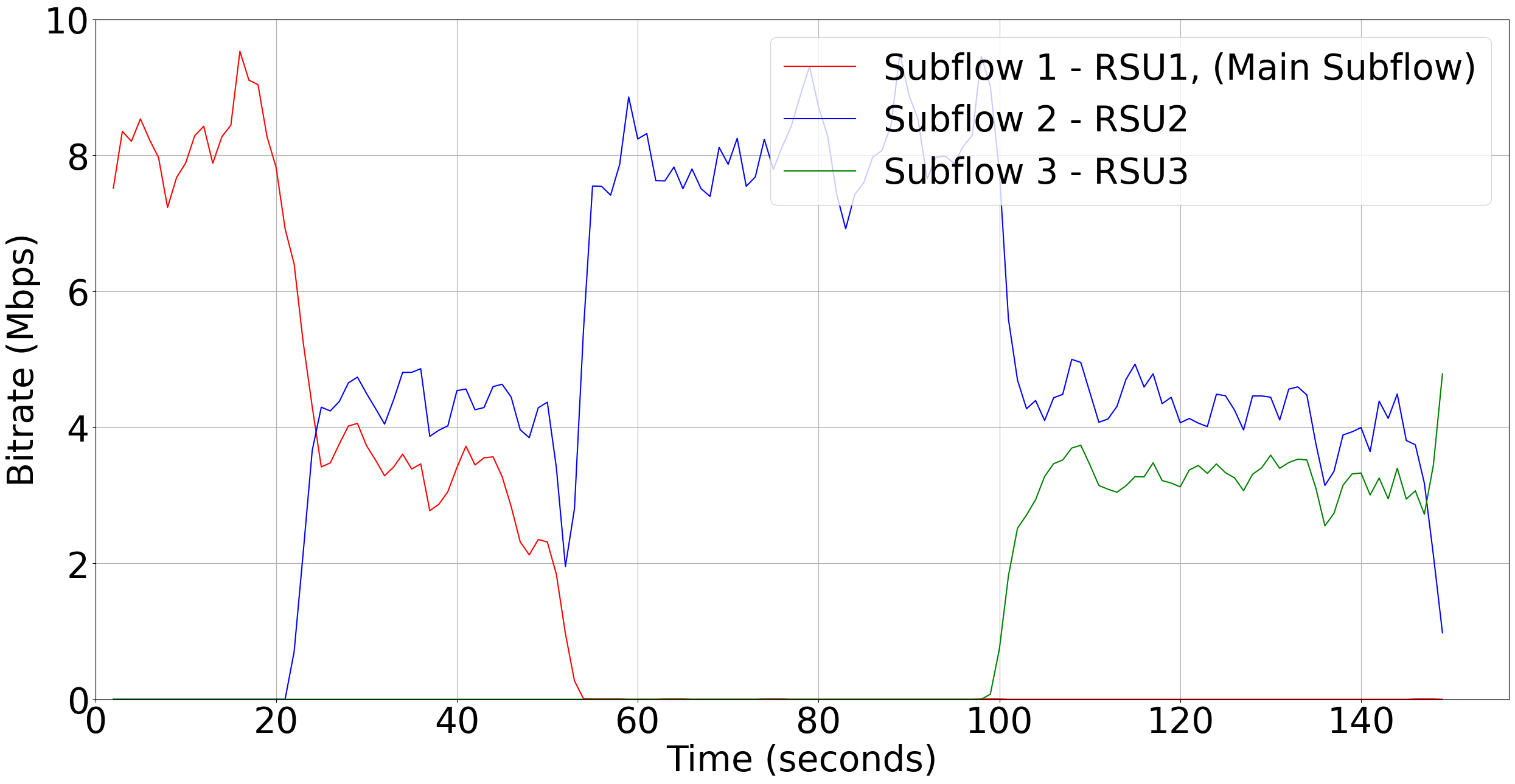}
        \caption{Bitrate.}
        \label{fig:bitrate}
    \end{subfigure}
    \hspace{0.2cm} 
    \begin{subfigure}[b]{0.4\textwidth}
        \centering
        \includegraphics[width=\textwidth]{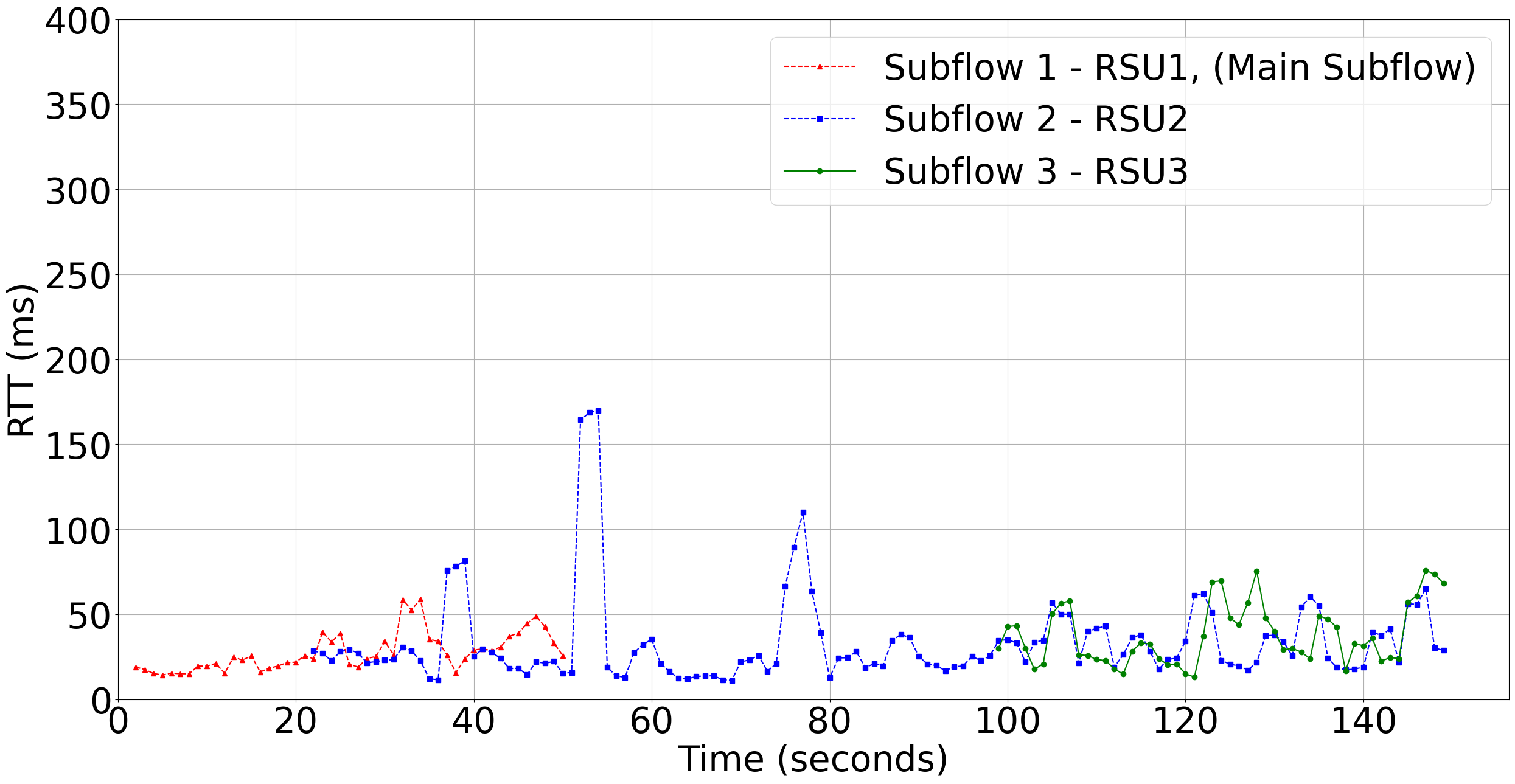}
        \caption{RTT.}
        \label{fig:rtt}
    \end{subfigure}
    \caption{MPTCP performance in the first scenario (no additional delay): a) bitrate and b) RTT values for the entire experimentation period.}
    \label{fig:SingleRat_Normal}
\end{figure}

Figure~\ref{fig:bitrate} presents the bitrate achieved by each MPTCP subflow during the experiment for the first scenario, \textit{i.e.} with no additional delay. To better understand the results achieved by the MPTCP, Fig.~\ref{fig:rtt} plots the measured RTT on each MPTCP subflow. The first observation goes to the correct use of all the available logical links between the OBU and the infrastructure: when a new RSU is in communication range, the subflow is created, and when a given RSU becomes unreachable, the respective MPTCP subflow is removed. However, the creation and removal of a MPTCP subflow does not imply a correct exploration of MPTCP: a subflow may be created but remains unused. Nevertheless, the results show that all the MPTCP subflows present considerable bitrate values.

A more detailed analysis shows that the first MPTCP subflow (denoted by the MPTCP as the main subflow), which is responsible for initiating the MPTCP connection, is established at the beginning of the experiment, and data is transmitted through it, reaching an average bitrate of 8.5 Mbps\footnote{IEEE 802.11p radios were configured for a maximum bitrate of 9 Mbps.}. When the second RSU becomes in range, an additional MPTCP subflow is created, leading to a reduction of the bitrate in the first subflow by approximately half. This occurs because the capacity of the OBU's radio interface is now shared between the two MPTCP subflows. When the connection with the first RSU is lost, we can observe an increase in the bitrate of the second MPTCP subflow justified by the fact that the radio interface is not being shared anymore between multiple logical links. The achieved bitrate is in line with the one achieved solely by the first MPTCP subflow. Finally, with the appearance of a third RSU, and consequently, a new logical link, the MPTCP explores both MPTCP subflows simultaneously.

Figure~\ref{fig:rtt} shows that the RTT values for all the active wireless links are very close. This justifies the almost equal distribution of the traffic over the simultaneous MPTCP subflows: the minRTT scheduling algorithm was used, resulting in a preference of paths with the lowest RTT. A more detailed analysis shows that, when MPTCP subflows 1 and 2 coexist, the average RTT is lower for the subflow 2 (blue curve), resulting in a slightly higher bitrate allocated to this subflow. The same occurs for the coexistence of MPTCP subflows 2 and 3.

\begin{figure}[ht]
    \centering
    \begin{subfigure}[b]{0.4\textwidth}
        \centering
        \includegraphics[width=\textwidth]{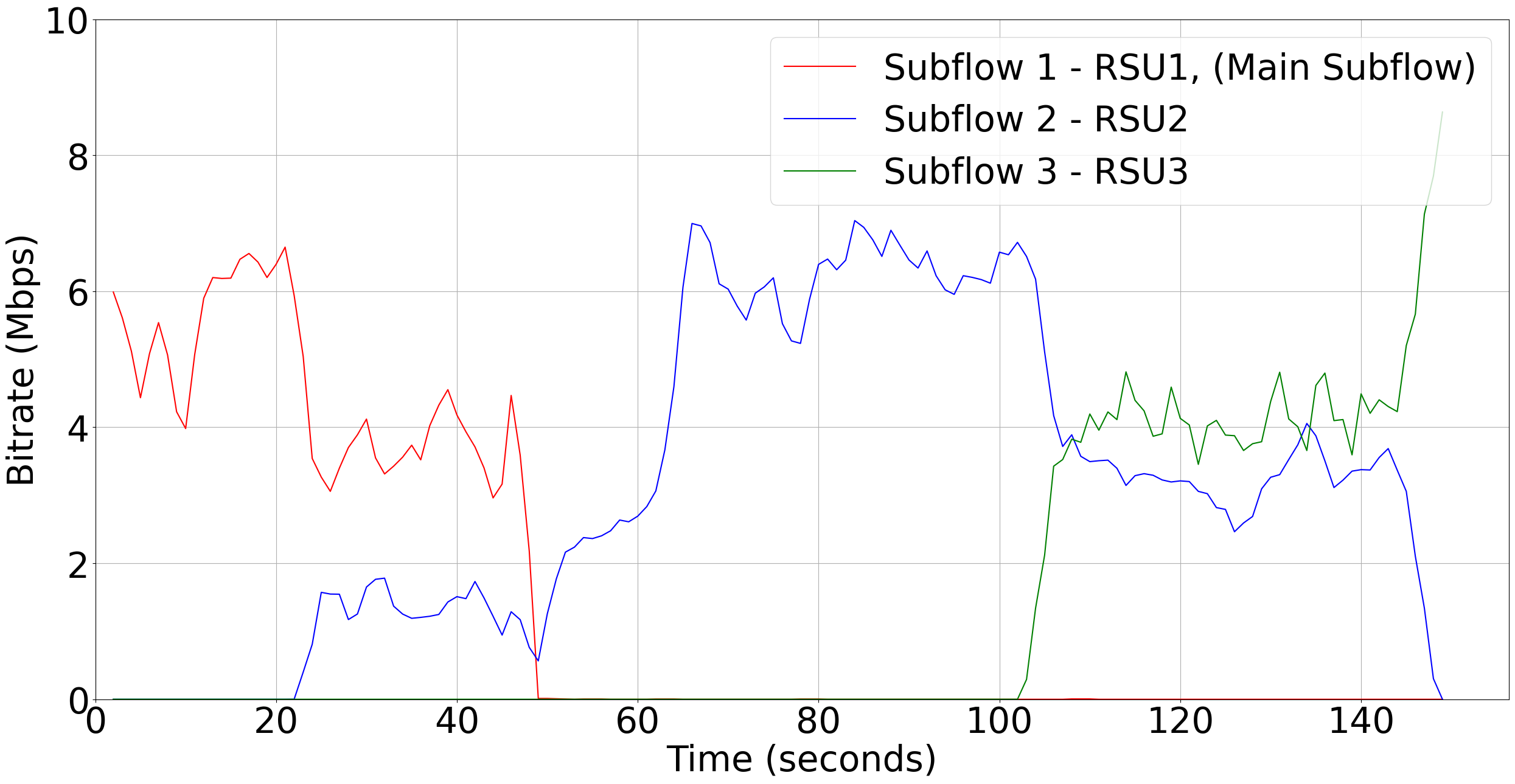}
        \caption{Bitrate.}
        \label{fig:bitrate2}
    \end{subfigure}
    \hspace{0.2cm} 
    \begin{subfigure}[b]{0.4\textwidth}
        \centering
        \includegraphics[width=\textwidth]{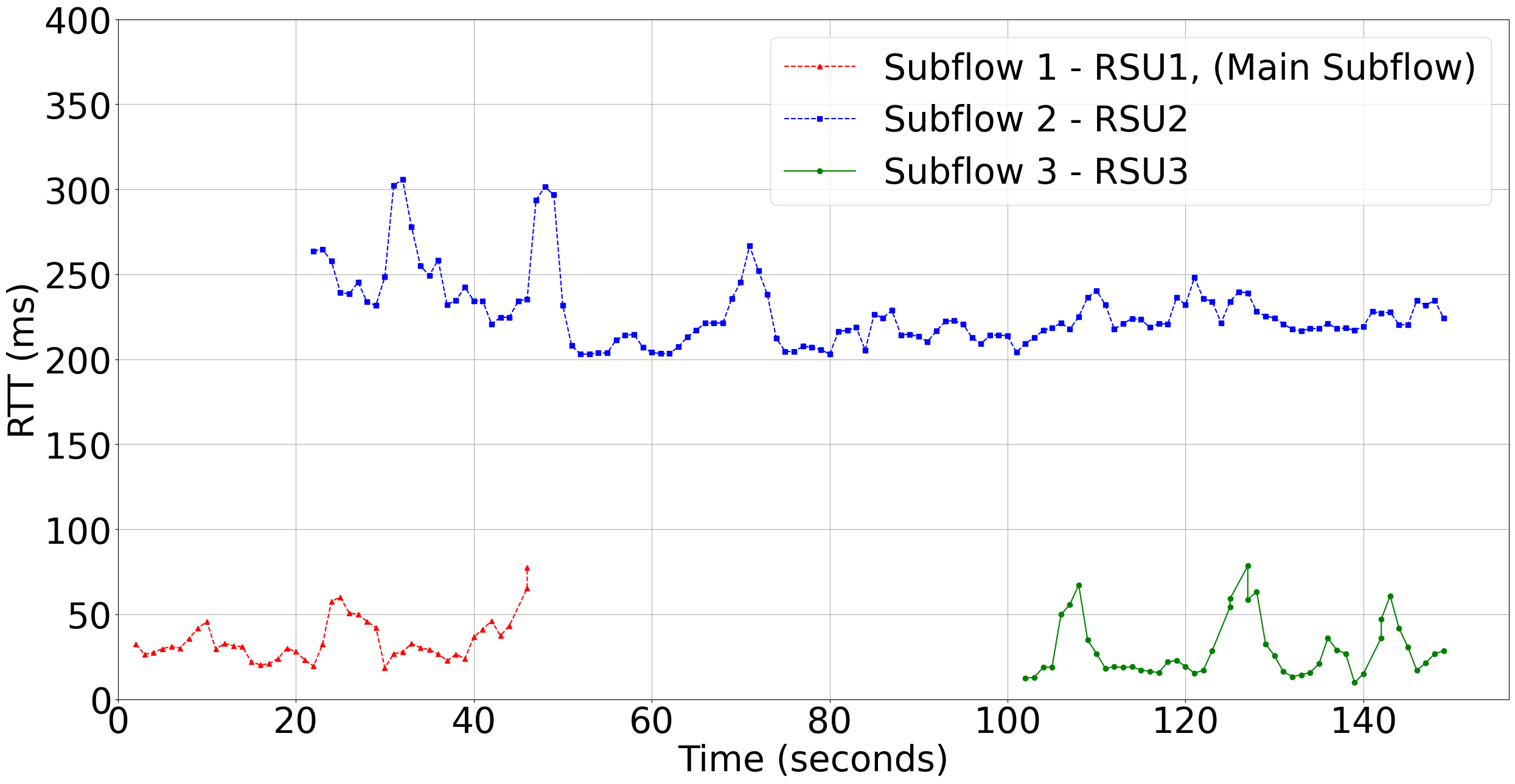}
        \caption{RTT.}
        \label{fig:rtt2}
    \end{subfigure}
    \caption{MPTCP performance in the second scenario (200 ms additional delay on the second link): a) bitrate and b) RTT values for the entire experimentation period.}
    \vspace{-3mm}
    \label{fig:SingleRat_Delay2nd}
\end{figure}

Figure~\ref{fig:SingleRat_Delay2nd} evaluates the MPTCP for the second scenario, when an additional delay is added to the second wireless logical link. In general, the behavior of the MPTCP is the same as observed before: MPTCP subflows are created and used when new RSUs become in range of the OBU, and removed otherwise. As for the maximum achieved bitrate on each MPTCP subflow, or as a result of the sum of all simultaneous MPTCP subflows, it is generally lower than the one observed before: approximately 6.2~Mbps when only RSU1 exists, 6~Mbps when subflows 1 and 2 coexist, 7~Mbps when only RSU2 exists, and approximately 8~Mbps when MPTCP subflows 2 and 3 coexist. Such result is justified by the increase in the channel occupation by other IEEE 802.11p stations, and not by the behavior of the MPTCP nor the additional delay introduced in the second wireless link.

Finally, with respect to the MPTCP packet scheduling algorithm and the impact of having highly asymmetric logical links over the same shared radio interface (with RTT values of 10 times more), we can see that, as expected, preference is given to subflows with lower RTT, more evident in the coexistence between subflows 1 and 2.

\section{Conclusions}\label{sec::5}

In mobile networks, the connectivity between the mobile nodes and the infrastructure is pivotal: without it, the mobile nodes may become isolated. To reduce such risk, mobile nodes are usually equipped with multiple RATs, complementing each other. Such configuration is explored by the MPTCP, where each MPTCP subflow is established over each RAT, allowing the simultaneous transmission of traffic through the different paths, with the purpose of increasing the performance of the traditional TCP. However, assuming that such multi-connectivity is always available is far from the reality. Thus, this pioneering work presents a different perspective on the use of MPTCP: can we use it in single-RAT environments?

With a vehicular network as use case, the results have shown that MPTCP can be effectively used when a single radio access technology is shared to create simultaneous communication paths between the mobile node and the infrastructure. Based on the quality of each wireless logical link, the MPTCP adjusts the transmission rate on each subflow, guaranteeing a proper operation with one or more active links. Such behavior provides a seamless handover process in vehicular networks, even when OBUs are equipped with a single radio access technology. Future work will extend the single-RAT configuration to include additional RATs to support more MPTCP subflows.

\bibliographystyle{IEEEtran}
\bibliography{references}

\end{document}